\documentclass[12pt]{article}
\usepackage{graphicx}

\newcommand\pubnumber{SNSN-XXX-YY}
\newcommand\pubdate{\today}

\def\napoli{University of Minnesota\\
116 Church ST. SE., Minneapolis, MN55455}
\def\support{\footnote{}}

\def\Title#1{\begin{center} {\Large #1 } \end{center}}
\def\Author#1{\begin{center}{ \sc #1} \end{center}}
\def\Address#1{\begin{center}{ \it #1} \end{center}}

\def\mpsipp{m_{\psi'\pi^-}}

\def\thetaks{\theta_{K^*}}
\def\cosks{\cos\thetaks}

\newcommand\pubblock{\rightline{\begin{tabular}{l} \pubnumber\\
         \pubdate  \end{tabular}}}
\newenvironment{Abstract}{\begin{quotation}  }{\end{quotation}}
\newenvironment{Presented}{\begin{quotation} \begin{center} 
             PRESENTED AT\end{center}\bigskip 
      \begin{center}\begin{large}}{\end{large}\end{center} \end{quotation}}
\def\Acknowledgements{\bigskip  \bigskip \begin{center} \begin{large}
             \bf ACKNOWLEDGEMENTS \end{large}\end{center}}

\begin{document}
\begin{titlepage}
\pubblock

\vfill
\Title{A Review of Heavy Exotic States}
\vfill
\Author{ Jianming Bian\support}
\Address{\napoli}
\vfill
\begin{Abstract}
A review of recent experimental developments concerning the $X$, $Y$ and $Z$ charmonium-like exotic states.
\end{Abstract}
\vfill
\begin{Presented}
XXXIV Physics in Collision Symposium \\
Bloomington, Indiana,  September 16--20, 2014
\end{Presented}
\vfill
\end{titlepage}
\def\thefootnote{\fnsymbol{footnote}}
\setcounter{footnote}{0}

\section{Introduction}

In the quark model, hadrons are dominantly bound states of $q\bar{q}$ (mesons) or $qqq$ (baryons), but QCD allows hadrons with $N_{quarks}\neq 2, 3$. These exotic states include glueballs (consist solely of gluon particles, with no quark), hybrids (with quarks and excited gluon), multi-quark states (with more than three quarks), and hadron molecules (bound states of two or more hadrons like the deuteron). There is a long history of searches for these exotic hadrons, however, no solid experimental evidence was found until recent breakthroughs in the charmonium region.

Below open-charm threshold, all charmonium states have been observed, and $c\bar{c}$ potential models describe the charmonium spectrum very well. However, many expected charmonium states are still missing above the open-charm threshold.  Since 2003, Belle, BaBar, BESIII and LHCb have made radical progress in studies of exotic heavy states  in this energy region by observing a number of new states above open-charm threshold which decay to charmonium and light hadrons. These states are so-called charmonium-like states or $XYZ$ particles. Although some of these particles could be interpreted as charmonium states, many of them have exotic properties that are not consistent with conventional mesons or baryons in the quark model.

The charmonium-like states are classified in three categories, $X$, $Y$ and $Z$ particles. $X$ particles are neutral and discovered  in $B$ decays and hadron machines, $Y$ particles are neutral vectors ($1^-$) observed in $e^+e^-$ colliders and $Z^\pm$ particles are charged quarkonium-like particles. In this talk, I will briefly review recent experimental developments about the $X$, $Y$ and $Z$ charmoniumlike exotic states.

\section{The $X(3872)$}

The first observed charmonium-like exotic state is the $X(3872)$.  It was discovered by Belle in 2003~\cite{belle_x3872} in the mode $\pi^+\pi^- J/\psi$ from $B^+\to K^+\pi^+\pi^- J/\psi$ decays and was subsequently confirmed by CDF~\cite{CDF_x3872}, D0~\cite{D0_x3872} and BaBar~\cite{babar_x3872}. The mass of $X(3872)$ is found to be close to the $D^0\bar{D}^{*0}$ threshold and the width is very narrow ($<$1.2 MeV). The mass value and the absence of a strong signal in the $\gamma\chi_{cJ}$ decay channel disagree with potential model expectations for a conventional charmonium state. Another feature of the $X(3872)$ that differs from conventional charmonium is that the decay branching ratio of $X(3872)$ to $\pi^+\pi^- J/\psi$ is comparable to $\pi^+\pi^-\pi^0 J/\psi$ \cite{delAmoSanchez:2010jr}, so the isospin mixing occurs on a large scale. Possible interpretations of $X(3872)$ include D- or P-wave charmonium, a hadronic molecule, a tetraquark and  molecule + charmonium mixture.

\begin{figure}[htbp]
\begin{center}
\includegraphics[width=8cm]{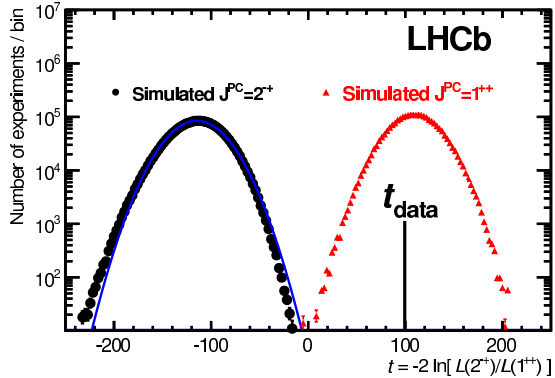}
\caption{Distribution of the test statistic $t$ for the simulated experiments  with $J^{PC}=2^{-+}$ and $\alpha=\hat{\alpha}$ (black circles on the left) and with $J^{PC}=1^{++}$ (red triangles on the right).  A Gaussian fit to the $2^{-+}$ distribution  is overlaid (blue solid line). The value of the test statistic for the data, $t_{\rm data}$, is shown by the solid vertical line.  }\label{fig:xjpc}
\end{center}
\end{figure}

BaBar and Belle observed $X(3872)$ decays to $\gamma J/\psi$, so $X(3872)$ should be a C-even state \cite{Aubert:2006aj,Bhardwaj:2011dj}. By analyzing angular distributions of $B^+\to $X(3872)$ K^+, X(3872)\to\pi^+\pi^- J/\psi$, CDF excluded all $J^{PC}$ hypotheses except $2^{-+}$ and $1^{++}$  \cite{Abulencia:2006ma}. A recent LHCb analysis based on angular correlations in the same channel determined $J^{PC}$ of $X(3872)$ to be $1^{++}$ unambiguously \cite{Aaij:2013zoa}, as shown in Figure~\ref{fig:xjpc}. This $J^{PC}$ value rules out the explanation of the $X(3872)$ meson as a conventional $\eta_{c2}(1^1D_2)$ state. The remaining possibility, $\chi_{c1}(2^3P_1)$ charmonium, is disfavored by the value of the $X(3872)$ mass.

Radiative decays of the $X(3872)$ to charmonium are important to check whether the $X(3872)$ is a conventional charmonium or an exotic state. $X(3872)\to\gamma J/\psi$ was measured by BaBar and Belle  \cite{Aubert:2006aj,Bhardwaj:2011dj}. Evidence for $X(3872)\to\gamma\psi(2S)$ was reported by BaBar \cite{Aubert:2008ae} (2008) and LHCb \cite{Aaij:2014ala} (2014). The statistical significances of $X(3872)\to\gamma\psi(2S)$ are $3.5\sigma$ for BaBar and $4.4\sigma$ for LHCb. For the ratio of $\mathcal{B}(X(3872)\to\gamma\psi(2S))$ to 
$\mathcal{B}(X(3872)\to\gamma J/\psi)$, 

BaBar obtained:

$$R=\frac{\mathcal{B}(X(3872)\to\gamma\psi(2S))}{\mathcal{B}(X(3872)\to\gamma J/\psi)}=3.4\pm1.4,$$

LHCb obtained:

$$R=\frac{\mathcal{B}(X(3872)\to\gamma\psi(2S))}{\mathcal{B}(X(3872)\to\gamma J/\psi)}=2.46\pm0.64\pm0.29,$$

Belle did not observe a significant signal for $X(3872)\to\gamma\psi(2S)$, so they set an upper limit of R $<$ 2.1 at the $90\%$ C.L \cite{Bhardwaj:2011dj}. Considering statistical and systematic errors, this upper limit does not contradict BaBar and LHCb results.

Averaging results from BaBar, Belle and LHCb, one can have:
$$\bar{R}=\frac{\mathcal{B}(X(3872)\to\gamma\psi(2S))}{\mathcal{B}(X(3872)\to\gamma J/\psi)}=2.31\pm0.57.$$

Since $R$ is expected to be very small for the pure molecule hypnosis, this experimental value does not support the pure $D^0\bar{D}^{0*}$ molecular interpretation of the $X(3872)$.  However, mixing of $D^0\bar{D}^{0*}$ molecule and charmonium could explain the enhanced $R$ value of $X(3872)$ \cite{Aaij:2014ala,Brambilla:2010cs}.

\section{$Y$-family states}
$Y$ states are a family of vectors $(J^{PC}=1^{--})$ observed in $e^+e^-$ colliders. In the ISR process $e^+e^-\to \gamma_{ISR} \pi^+\pi^- J/\psi$ , the BaBar experiment observed the $Y(4260)$ state in $\pi^+\pi^- J/\psi$ \cite{Aubert:2005rm}. This state was confirmed by the CLEO \cite{He:2006kg} and Belle \cite{Yuan:2007sj} experiments in the same channel. It was found that the properties of $Y(4260)$ are different from conventional $1^{--}$ charmonium, because it strongly couples to $\pi^+\pi^-J/\psi$ and has no significant enhancement in open charm production. After the discovery of $Y(4260)$, several $Y$ states were observed by Belle and BaBar in $\pi^+\pi^-J/\psi$ ($Y(4008)$)\cite{Yuan:2007sj}, $\pi^+\pi^-\psi(2S)$ ($Y(4360)$, $Y(4660)$)\cite{Aubert:2007zz, Wang:2007ea}  and $\Lambda_c\bar{\Lambda_c}$ (Y(4630)) \cite{Pakhlova:2008vn} from $e^+e^-$ ISR processes. Among these states, $Y(4260)$, $Y(4360)$ and $Y(4660)$ were solidly confirmed by Belle and BaBar with large data samples \cite{Aubert:2007zz,Liu:2013dau,Lees:2012pv,Wang:2014hta}. An attempt to fit the $\pi^+\pi^- J/\psi$ spectrum with coherent BW functions for $Y(4260)$, $Y(4360)$ and $Y(4660)$ was done by Belle \cite{Wang:2014hta}. A summary of observed $Y$ states is listed in Table~\ref{tab:sumy}.

\begin{table}[h]
\begin{center}
\begin{tabular}{|c|c|c|c|c|}  
\hline
State &  Mass (MeV/$c^2$) & Width (MeV/$c^2$) &  Decay mode & Experiment\\ \hline

$Y(4008)$ &  $4008^{+121}_{-49}$ & $226\pm97$ &  $\pi^+\pi^{-}J/\psi$ & Belle\\ \hline

$Y(4260)$ &  $4250\pm9$ & $108\pm12$ &  $\pi^+\pi^-J/\psi$ & BaBar\\
&&&$\pi^0\pi^0J/\psi$&CLEO\\ 
&&&$K^+K^-J/\psi$&Belle\\\hline

$Y(4360)$ &  $4361\pm13$ & $74\pm18$ &  $\pi^+\pi^-\psi(2S)$ & Belle\\
&&&$K^+K^-J/\psi$&BaBar\\\hline

$Y(4630)$ &  $4634^{+9}_{-11}$ & $92^{+41}_{-32}$ &  $\Lambda_c\bar{\Lambda_c}$ & Belle\\ \hline

$Y(4660)$ &  $4664\pm12$ & $48\pm15$ &  $\pi^+\pi^-\psi(2S)$ & Belle\\
&&&&BaBar\\\hline

\end{tabular}
\caption{List of $Y$-family states.}
\label{tab:sumy}
\end{center}
\end{table}

Between 4 and 4.7 GeV, at most five $1^{--}$ states are expected in the charmonium family (3S, 2D, 4S, 3D, 5S). However, including $Y$ states, seven particles are observed. This is another hint that some of these $Y$ states should be exotic states. Possible explanations include hybrids, Molecular states, Hadrocharmonium, threshold effect and FSI effects. Although the exact nature of the $Y$ family is still unclear, we know that $Y(4260)$, $Y(4360)$ and $Y(4660)$ are similar and narrow.

\begin{figure}[htbp]
\begin{center}
\includegraphics[width=10cm]{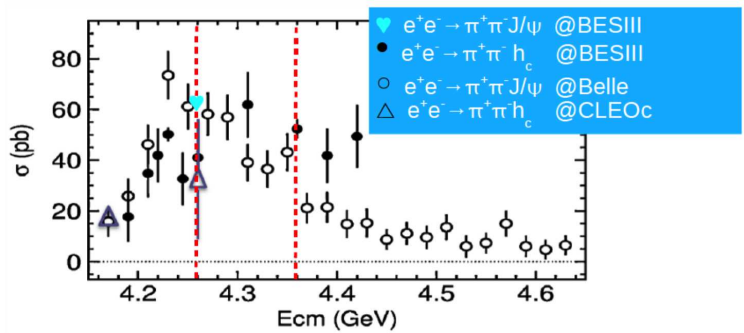}
\caption{Cross sections of charmonium productions around $Y$ states region} \label{fig:xsec}
\end{center}
\end{figure}

Another path towards understanding the $Y$-family is measuring cross sections of charmonium productions around the $Y$-family region. Cross sections of $\pi^+\pi^-J/\psi$ and $\pi^+\pi^- h_c$ measured by Belle, BESIII and CLEOc are shown in Figure~\ref{fig:xsec}. One can see that different decay modes have similar cross sections but the line shapes seem to be different in these modes. This difference could be explained by threshold or interference effects.

\begin{figure}[htbp]
\begin{center}
\includegraphics[width=8cm]{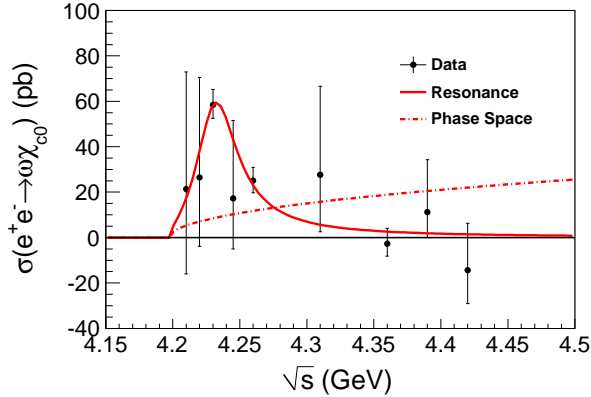}
\caption{Fit to $\sigma(e^+e^-\to\omega\chi_{c0})$ with a resonance (solid curve), or a phase space term (dot-dashed curve). Dots with error bars are the dressed cross sections. The uncertainties are statistical only. }\label{fig:xsecomegachic}
\end{center}
\end{figure}

To study the coupling between the $Y(4260)$ and the $\omega \chi_{cJ}$ channels, BESIII recently measured the cross section of $e^+e^-\to\omega \chi_{c0}$ from 4.21 to 4.42 GeV \cite{Ablikim:2014qwy}. As shown in Figure~\ref{fig:xsecomegachic}, $\omega \chi_{c0}$ are observed at 4230 MeV and 4260 MeV. One interesting thing is that the signal of $\omega \chi_{c0}$ does not arise from the decays of the $Y(4260)$ and can be fit with a resonance peaking at $4230\pm8$ MeV. By comparing with $\omega\chi_{c0}$ results, the measured
$e^+e^-\to \omega\chi_{c1,2}$ cross sections are found to be much lower than the predictions in Ref.~\cite{Ratio}. More statistics and energy points are needed to understand whether the enhancement of $\omega \chi_{c0}$ is caused by threshold effects, interference or a new resonance.

BESIII reported the observation of $e^+e^-\to\gamma X(3872), X(3872)\to\pi^+\pi^-J/\psi$ with data collected from 4.009 to 4.420 GeV with a statistical significance of $6.3\sigma$ \cite{Ablikim:2013dyn}. By fitting the energy-dependent cross sections  with different hypotheses, they found that the $Y(4260)$ resonance describes the data better than a linear continuum or an E1 transition phase space shape. The ratio $R(\mathcal{B}(e^+e^-\to\gamma X(3872))/\mathcal{B}(e^+e^-\to\pi^+\pi^-J/\psi))$ is then determined to be $\sim11\%$, which represents a large transition rate. Together with the observation of $Z_c$(3900) in $Y(4260)$ decays  \cite{Ablikim:2013mio}, this result may indicate the commonality in the nature of the exotics states $X(3872)$, $Y(4260)$, and $Z_c(3900)$.

\section{Z particles}
Considering the difficulties in interpreting $X$ and $Y$ states, searching for the charged charmonium-like states, $Z$ particles, is the most direct way to identify exotic hadrons. Decaying to charmonium demonstrates that a $Z$ particle has a $c\bar{c}$ pair. The electric charge indicates that it must have two or more light quarks. Therefore, a $Z$ particle consists of at least four quarks, which is a clear signature for an exotic hadronic state. Promising channels to look for $Z_c$ includes $\pi^{\pm}J/\psi$, $\pi^{\pm} h_c$, $\pi^{\pm} \psi(2S)$ $\pi^{\pm}\chi_{cJ}$, etc.

\begin{figure}[htbp]
 \includegraphics[height=5cm]{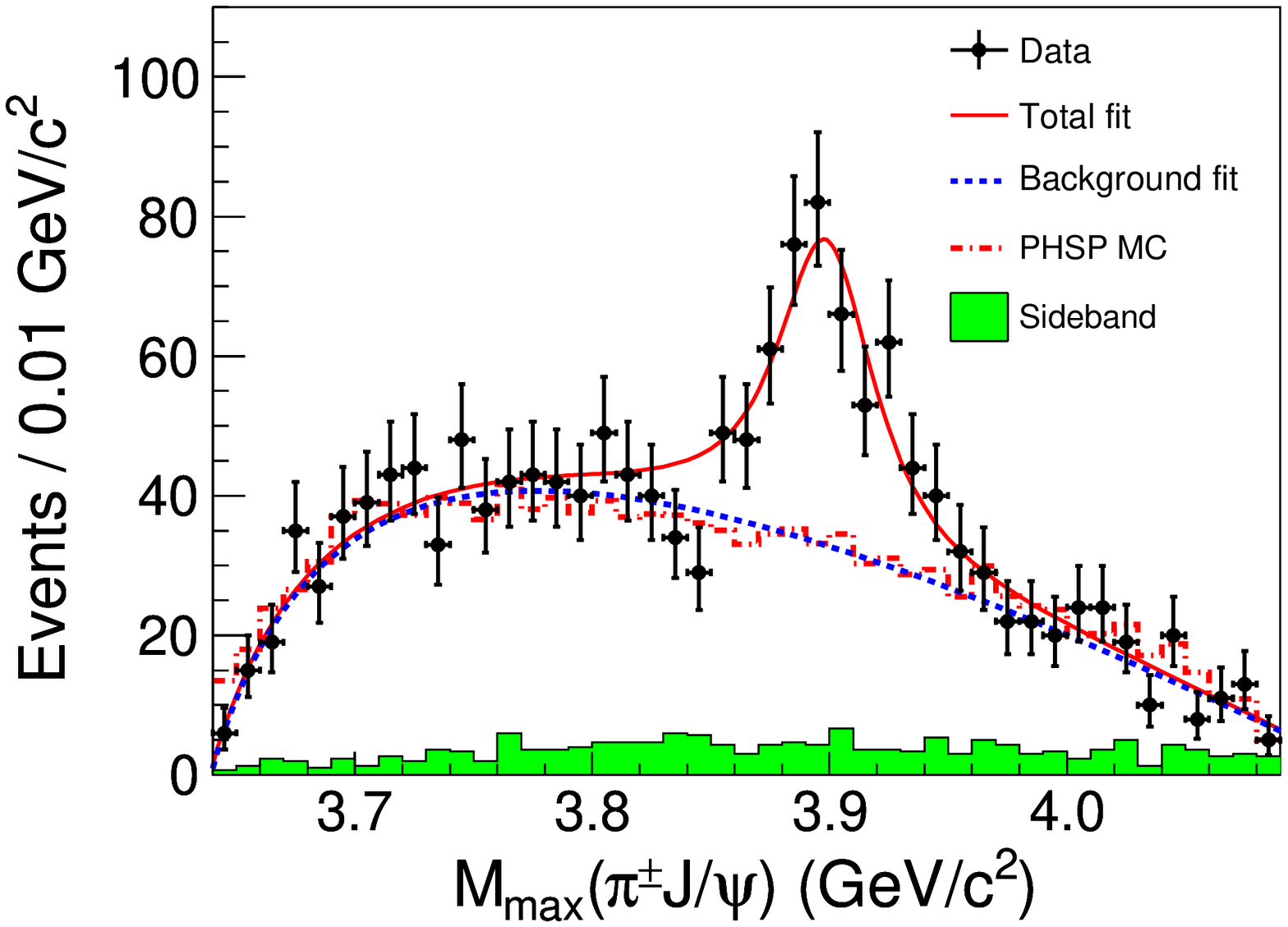}
 \includegraphics[height=5cm]{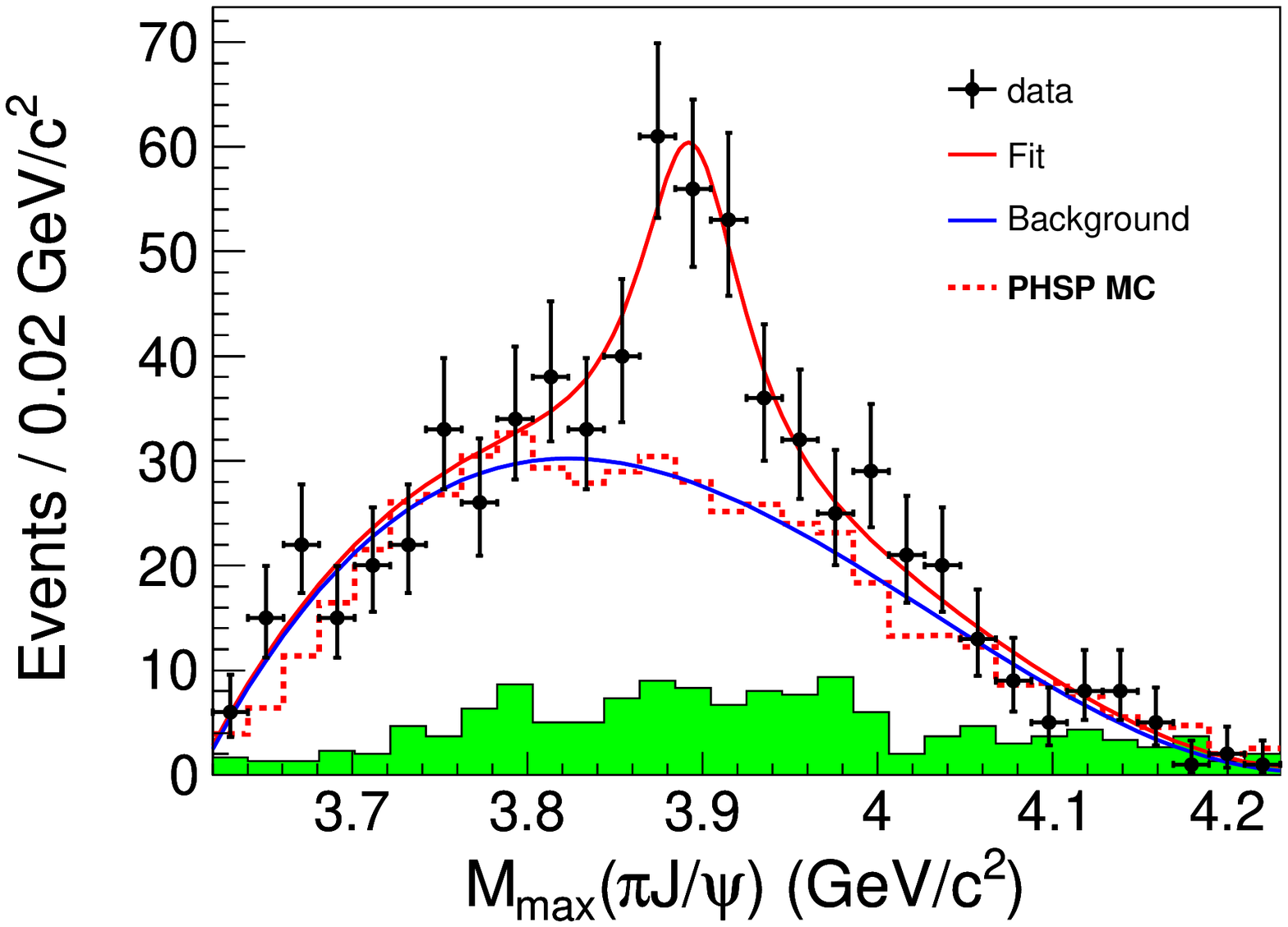}
 \caption{Fits to the $M_{\mathrm{max}}(\pi J/\psi)$ spectrum for the observation of $Z_c(3900)^\pm$, points with error bars are data, the
curves are the best fit, the dashed histograms are the phase space
distributions and the shaded histograms are the non-$\pi^+\pi^- J/\psi$
background estimated from the normalized $J/\psi$ sidebands(left: BESIII,
right: Belle). }
\label{fig:zcfit}
\end{figure}

The first confirmed charged charmornium-like state is $Z_c(3900)^\pm$. It was observed in $\pi^+\pi^-J/\psi$ by BESIII \cite{Ablikim:2013mio} ($E_{cm}$=4.17 GeV, 525 pb$^{-1}$), Belle \cite{Liu:2013dau} (ISR from 3.8 to 5.5 GeV, 967 fb$^{-1}$) and confirmed in CLEO-c data \cite{Xiao:2013iha}(NWU group, $E_{cm}$=4.17 GeV, 586pb$^{-1}$). The mass was determined to be $3899.0\pm3.6\pm4.9\rm{MeV}/c^2$ (BESIII), $3894.5\pm6.6\pm4.5\rm{MeV}/c^2$ (Belle) and $3886\pm4\pm2\rm{MeV}/c^2$ (CLEO-c data) and the width was determined to be $46\pm10\pm20\rm{MeV}/c^2$ (BESIII), $63\pm24\pm26\rm{MeV}/c^2$ (Belle) and $37\pm4\pm8\rm{MeV}/c^2$ (CLEO-c data), respectively. Fits to the $M_{\mathrm{max}}(\pi J/\psi)$ spectrum from BESIII and Belle can be found in Figure~\ref{fig:zcfit}. The neutral partner of $Z_c^{\pm}(3900)^\pm$, $Z_c(3900)^{0}$, is observed in $\pi^0\pi^0J/\psi$ by BESIII at $E_cm$=4230, 4260, 4360MeV \cite{beszc0}.  As discussed above, $Z_c(3900)$ contains at least four quarks. The mass of $Z_c$(3900) is close to the $D\bar{D}^*$ threshold, so interpretations of $Z_c$(3900) include a molecular state, a tetraquark, hadrocharmonium or threshold effects.

By studying $e^+e^-\to\pi^{\pm}(D\bar{D}^*)$ at 4.26 GeV, BESIII observed an enhancement in the $D\bar{D}^*$ invariant mass spectrum\cite{Ablikim:2013xfr}. Breit-Wigner resonance parameters are extracted to be $3883.9\pm1.5\pm4.2 \rm{MeV}/c^2$ for the mass and $24.8\pm3.3\pm11.0 \rm{MeV}/c^2$ for the width. This structure is named $Z_c$(3885). The angular distribution of the emitted $\pi$ in the $Z_c$(3885) system indicates that $J^p$ of  $Z_c$(3885) favors $1^+$. To determine whether $Z_c$(3885) and $Z_c$(3900) are the same state needs the study of mass/width differences between these two particles and the determination of $Z_c$(3900) $J^{P}$. Assuming the $Z_c$(3885) and $Z_c$(3900) are the same state, $\Gamma(D\bar{D}^*)/\Gamma(\pi^{\pm}J/\psi)$ can be determined to be $6.2\pm2.9$, which means a much larger non-$D\bar{D}$ branching ratio compared to conventional charmonium above the open charm threshold.

\begin{figure}[htbp]
\begin{center}
\includegraphics[width=8cm]{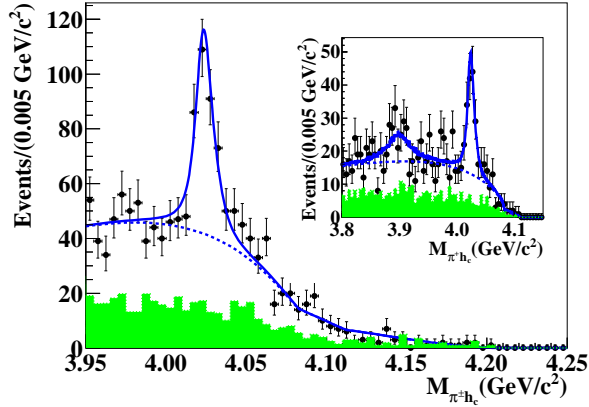}
\caption{Sum of the simultaneous fits to the $M_{\pi^{\pm} h_c}$
distributions at 4.23~GeV, 4.26~GeV, and 4.36~GeV as described in
the text; the inset shows the sum of the simultaneous fit to the
$M_{\pi^+ h_c}$ distributions at 4.23~GeV and 4.26~GeV with $Z_c$(3900)
and $Z_c$(4020). Dots with error bars are data; shaded histograms are
normalized sideband background; the solid curves show the total
fit, and the dotted curves the backgrounds from the fit.}\label{fig:zcp}
\end{center}
\end{figure}

BESIII observed another charged charmonium-like state $Z_c^{\pm}$(4020) in $e^+e^-\to\pi^+\pi^- h_c(1P)$ at CM energies between 3.90 and 4.42 GeV\cite{Ablikim:2013wzq} . As shown in Figure~\ref{fig:zcp}, a narrow resonance is discovered in the $\pi^{\pm}h_c$ spectrum and resonance parameters are determined to be $M = 4022.9\pm0.8\pm2.7 \rm{MeV}/c^2$ and $\Gamma = 7.9\pm2.7\pm2.6 \rm{MeV}/c^2$. The statistical significance of $Z_c^{\pm}$(4020) is $8.9\sigma$. In addition, the signal of $Z_c$(3900) is not significant in the $\pi^{\pm}h_c$ spectrum. A fit with both $Z_c$(3900) and $Z_c$(4020) to the data shows that the statistical significance of  $Z_c$(3900) is $2.1\sigma$. Because the spins of $c$ and $\bar{c}$ are in the same direction for $J/\psi$ and  are in different directions for $h_c$, a further study of $Z_c$(3900) and $Z_c$(4020) decaying to $\pi^{\pm} J/\psi$ and $\pi^{\pm} h_c$ can provide information about whether the $c$ and $\bar{c}$ in $Z_c$(3900) and $Z_c$(4020) are bounded tightly. As in the $Z_c$(3900) case, BESII also observed the neutral partner of $Z_c^{\pm}$(4020), $Z_c^{0}$(4020), in $\pi^0\pi^0 h_c$ with more than $5\sigma$ \cite{Ablikim:2014dxl} .

BESIII also studied the process $e^+e^-\to D^*\bar{D}^*)^{\pm}\pi^{\mp}$ at 4.26 GeV with a 827 pb$^{-1}$ data sample \cite{Ablikim:2013emm}. A structure near the $D^*\bar{D}^*$ threshold in the $\pi^{\pm}$ recoil mass spectrum is observed, which is named as the $Z_c$(4025). The mass and width  are determined to be $(4026.3\pm2.6\pm3.7) \rm{MeV}/c^2$ and $(24.8\pm5.6\pm7.7) \rm{MeV}/c^2$, respectively. Because the $Z_c$(4020) is also close to the $D^*\bar{D}^*$ threshold and has similar width, $Z_c$(4020) and $Z_c$(4025) may be the same state. Assuming Zc(4025) is Zc(4020), one can obtain  $\Gamma(D^*\bar{D}^*)/\Gamma(\pi^{\pm} h_c) = 12\pm5$, which is a large non-$D\bar{D}$ coupling.

Before the observation of $Z_c$, Belle observed two charged bottomonium-like resonances, $Z_b$(10610)$^{\pm}$ and $Z_b$(10650)$^{\pm}$, in $\Upsilon$(5S) decays using a data sample of 121.4 fb$^{-1}$ \cite{Belle:2011aa} . They observed these two resonances in both $\pi^{\pm}\Upsilon(nS)$ and $\pi^{\pm}h_b(mP)$ spectra. Masses (widths) are determined to be $10607.2\pm2.0 \rm{MeV}/c^2$ ($7.9\pm2.6 \rm{MeV}/c^2$) for $Z_b$(10610)$^{\pm}$ and $10652.2\pm1.5 \rm{MeV}/c^2$ ($11.5\pm2.2 \rm{MeV}/c^2$) for $Z_b$(10610)$^{\pm}$. $J^P$ are $1^+$ for both states. Since the $Z_b$(10610) is close to the threshold of $B\bar{B}^*$ and the $Z_b$(10650) is close to the threshold of $B^*\bar{B}^*$, one may naturally think that they are heavy flavor partners of $Z_c$(3900) and $Z_c$(4020). 

\begin{figure}[htbp]
 \includegraphics[height=5cm]{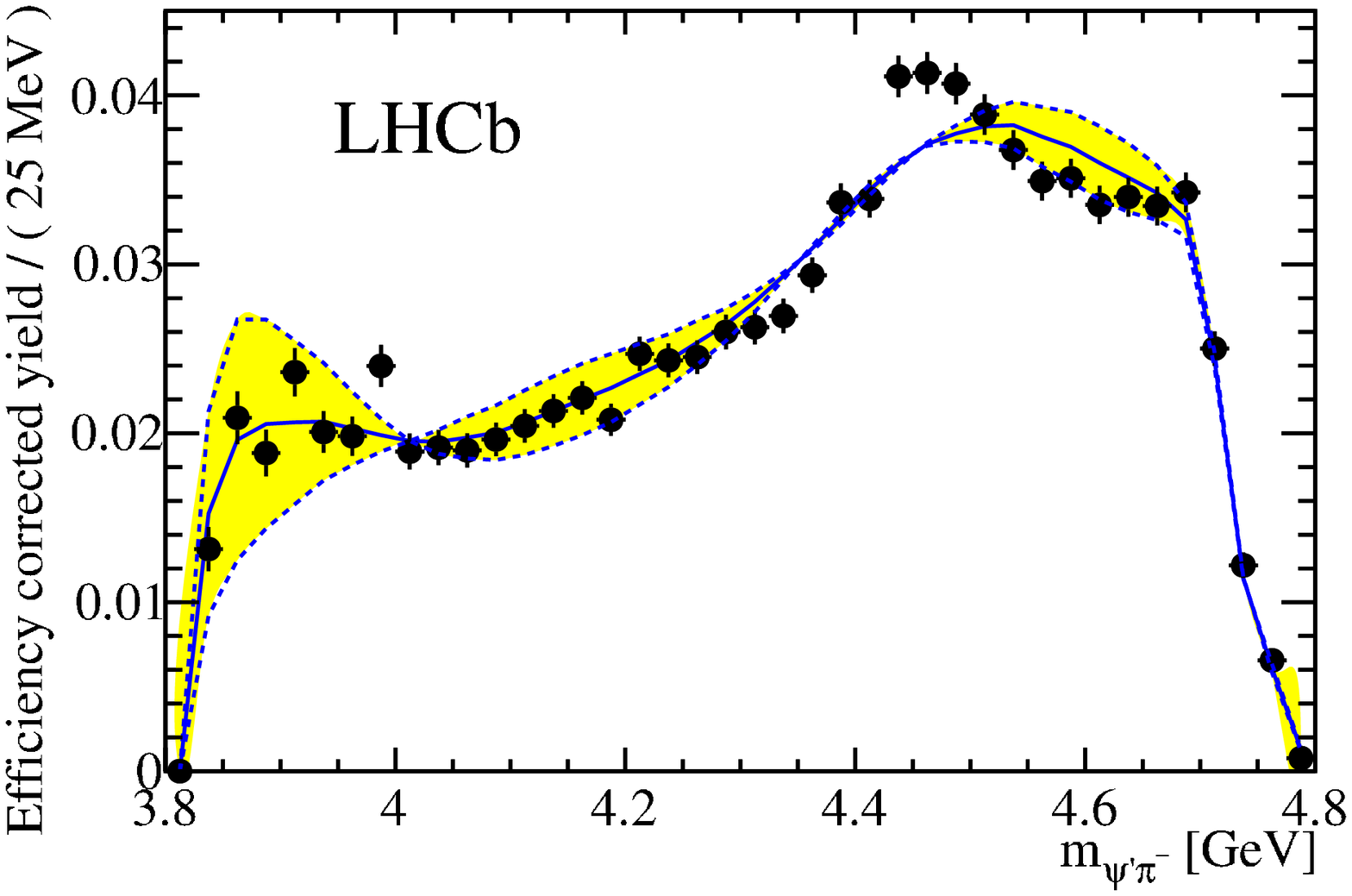}
 \includegraphics[height=5cm]{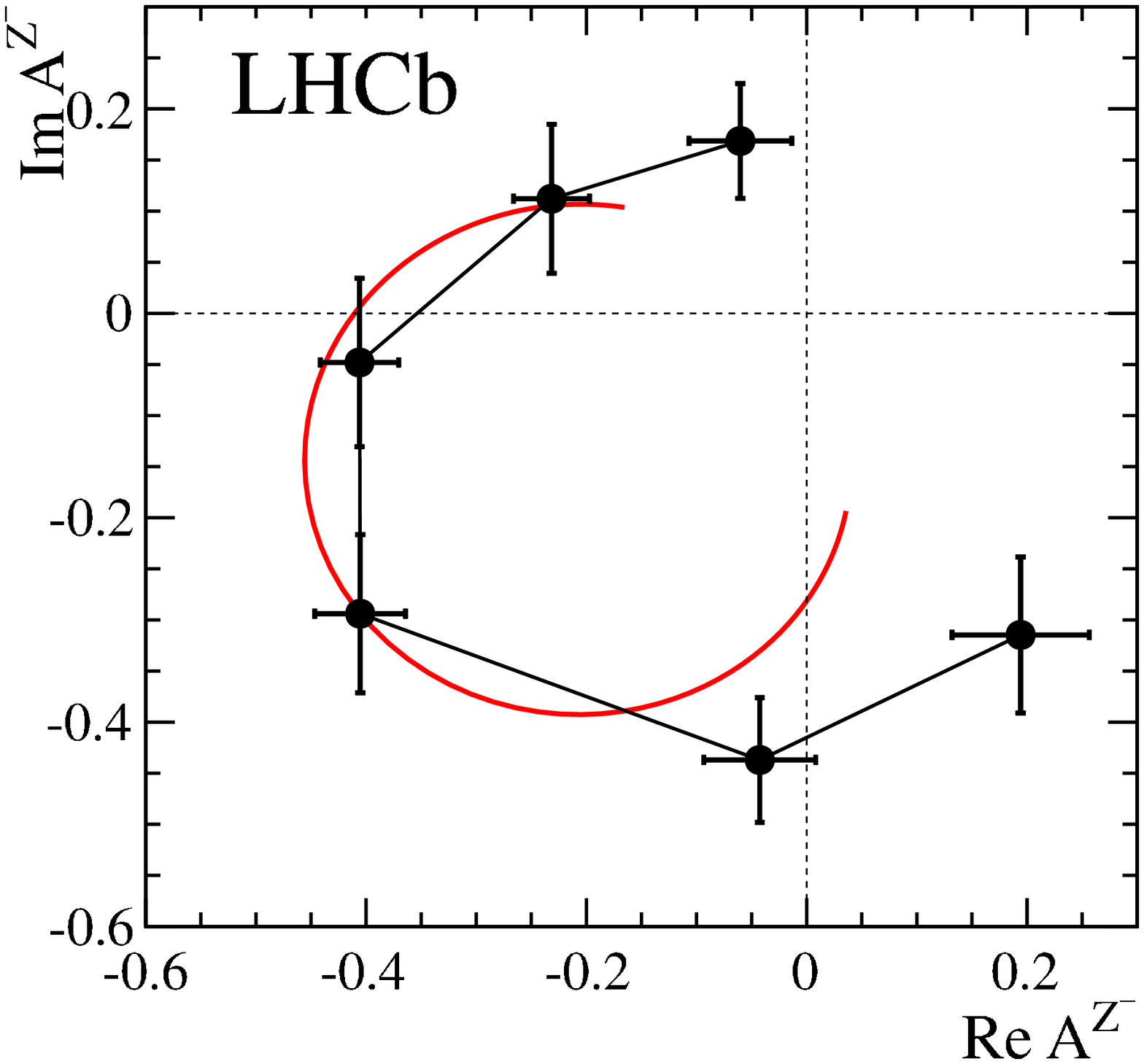}
 \caption{
 Left: Background-subtracted and efficiency-corrected $\mpsipp$ distribution (black data points), superimposed with the reflections of $\cosks$ moments up to order four allowing for $J(K^*)\le 2$  (blue line) and their correlated statistical uncertainty  (yellow band bounded by blue dashed lines). The distributions have been normalized to unity.;  Right: Fitted values of the $Z(4430)^-$ amplitude in six $\mpsipp^2$ bins,  shown in an Argand diagram (connected points with the error bars, $\mpsipp^2$ increases counterclockwise). The red curve is the prediction from the Breit-Wigner formula with a resonance mass (width) of 4475 (172) MeV and magnitude scaled to intersect the bin with the largest magnitude centered at (4477 MeV)$^2$. Units are arbitrary.  The phase convention assumes the helicity-zero $K^*(892)$ amplitude to be real.
 }
\label{fig:z4430fit}
\end{figure}

The Belle collaboration first reported evidence for the charged charmonium-like particle $Z(4430)^-$ in the $\pi^-\psi(2S)$ invariant mass spectrum in $B^0\to\psi(2S)K^-\pi^-$. The report mass and width are $M=(4433\pm4\pm2) \rm{MeV}/c2$ and $\Gamma= 45^{+18+30}_{-13-13} \rm{MeV}/c^2$ \cite{Choi:2007wga}. Although this result was confirmed by Belle's later data \cite{Chilikin:2013tch} , BaBar's data did not confirm the Belle's observation \cite{Aubert:2008aa}. Therefore, the existence of $Z(4430)^-$ has been doubted for a long time. By doing amplitude analyses using a much larger sample, the LHCb experiment recently confirmed Belle results\cite{Aaij:2014jqa}. The significance of $Z(4430)^-$ is greater than $13.9\sigma$. The mass and width are determined to be $4475\pm7\,{_{-25}^{+15}} ~\rm{MeV}/c^2$ and $172\pm13\,{_{-34}^{+37}}~\rm{MeV}/c^2$, and the spin-parity is determined to be $1^+$. In addition, LHCb also fits the real and imaginary parts of the $Z(4430)^-$ amplitude in six $\mpsipp^2$ bins around the $Z(4430)^-$ mass region. The Argand diagram is consistent with the behavior of a Breit-Wigner resonance, as shown in Figure~\ref{fig:z4430fit}. After the confirmation of $Z(4430)^-$ by LHCb, the Belle experiment recently observed a similar charmonium-like particle $Z_c(4200)$ in $B^0\to J/\psi K^-\pi^-$ with a significance of 6.2 $\sigma$ \cite{Chilikin:2014bkk}. A list of confirmed $Z_c$ states can be found in Table~\ref{tab:sumz}.

\begin{table}[h]
\begin{center}
\begin{tabular}{|c|c|c|c|c|c|}  
\hline
State &  Mass (MeV/$c^2$) & Width (MeV/$c^2$) &  Decay mode & $J^P$ & Experiment\\ \hline

$Z_c(3900)^{\pm}$ &  $3888.6\pm2.7$ & $34.7\pm6.6$ &  $\pi^{\pm}J/\psi~(D\bar{D}^*)$ & $1^+$ & BESIII\\
&&&&&Belle\\
&&&&&CLEO-c data\\\hline
$Z_c(4020)^{\pm}$ &  $4023.8\pm2.1$ & $7.9\pm3.8$ &  $\pi^{\pm} h_c~(D^*\bar{D}^*)$ & $1^?$ & BESIII\\ \hline
$Z_c(4430)^{\pm}$ &  $4478\pm21$ & $181\pm33$ &  $\pi^{\pm}\psi(2S)$ & $1^+$ & Belle\\ 
&&&&&LHCb\\ \hline

\end{tabular}
\caption{List of confirmed $Z_c$ states}
\label{tab:sumz}
\end{center}
\end{table}

\newpage

\section{Summary}
In summary, exotic heavy states, $XYZ$, are found both in charmonium and bottomonium sectors in the last decade. Many of them do not fit into the naive expectation of the conventional quarkonium spectrum. With large data samples in the charmonium region, $J^{PC}$ of $X(3872)$ has been determined and Y(4260),Y(4360) and Y(4660) have been confirmed.
Recent observations and confirmations of the charged charmonium states, $Z_c(3900)$, $Z_c(4020)$ and $Z(4430)^-$, may indicate that exotic states with four or more quarks have been observed. In addition to current data samples and analyses, one can expect LHCb, BelleII, BESIII and PANDA to produce more data in both  thecharmonium and bottomonium region to further the study of  heavy exotics states.

\Acknowledgements
I am grateful to the organizers of PIC2014 for the invitation. I appreciate my BESIII and UMN colleagues for helpful discussions.

\end{document}